\documentstyle[prd,aps,epsfig,amssymb]{revtex}
\let\jnfont=\rm
\def\NPB#1,{{\jnfont Nucl.\ Phys.\ B }{\bf #1},}
\def\PLB#1,{{\jnfont Phys.\ Lett.\ B }{\bf #1},}
\def\EPJC#1,{{\jnfont Euro.\ Phys.\ J.\ C }{\bf #1},}
\def\PRD#1,{{\jnfont Phys.\ Rev.\ D }{\bf #1},}
\def\PRL#1,{{\jnfont Phys.\ Rev.\ Lett.\ }{\bf #1},}
\def\MPLA#1,{{\jnfont Mod.\ Phys.\ Lett.\ A }{\bf #1},}
\def\JPG#1,{{\jnfont J.\ Phys.\ G}{\bf #1},}
\def\CTP#1,{{\jnfont Commun.\ Theor.\ Phys.\ }{\bf #1},}

\def\p_slash{\not{\hbox{\kern-2.1pt $p$}}}
\def\k_slash{\not{\hbox{\kern-2.1pt $k$}}}
\begin{document}
\draft
\preprint{}

\title{Studying top quark decay into the polarized W-boson in the TC2 model }

\author{Xuelei Wang$^{a,b}$, Qiaoli Zhang$^{b}$, Qingpeng Qiao$^{b}$\\
{\small a: CCAST(World Laboratory) P.O. BOX 8730. B.J.
100080P.R.China}\\ {\small b: College of Physics and Information
Engineering,}\\ \small{Henan Normal University, Xinxiang, Henan,
453007. P.R.China}
\thanks{E-mail:wangxuelei63@hotmail.com}
\thanks{Mailing address} }
\maketitle

\begin{abstract}

We study the decay mode of top quark decaying into Wb in the TC2
model where the top quark is distinguished from other fermions by
participating in a strong interaction. We find that the TC2
correction to the decay width $\Gamma (t \to b W) $ is generally
several percent and maximum value can reach $8\%$ for the
currently allowed parameters. The magnitude of such correction is
comparable with QCD correction and larger than that of minimal
supersymmetric model. Such correction might be observable in the
future colliders. We also study the TC2 correction to the
branching ratio of top quark decay into the polarized W bosons and
find the correction is below $ 1 \% $. After considering the TC2
correction, we find that our theoretical predictions about the
decay branching ratio are also consistent with the experimental
data.

\end{abstract}

\section{Introduction}

While the light quarks and leptons can be regarded as perturbative
spectators to the electroweak symmetry breaking (EWSB), the
massive top quark with a mass of EWSB scale suggests that top
quark is potentially enjoying a more intimate role in the flavor
dynamics and/or horizontal symmetry breaking. A potential
implication of this is the possibility that there exist extra
interactions for top quark, which distinguishs top quark from
other fermions of the standard model (SM) at the electroweak
scale. If this is true, the top quark physics will be much richer
than that of the SM and possible large deviations of top quark
properties from the SM predictions are expected \cite{sensitive}.
Detailed study about such new physics effects may reveal  useful
information about the underlying interactions and the mechanism of
EWSB. Such study is essential when one considers the advancement
in experiments where the forthcoming CERN Large Hadron Collider
(LHC) and the planned Next-generation Linear Collider (NLC) will
sever as top quark factories and thus make the precise
measurements of top quark properties possible~\cite{review}. In
this paper, we restrict our discussion in the framework of the
topcolor-assisted technicolor model
(TC2)\cite{tc2-Hill,tc2-Lane,tc2-2}, In this model, the third
generation is singled out to participate in a special interaction
called topcolor interaction. Such an interaction will cause the
top quark condensation which can partially contribute to EWSB and
also provide main part of top quark mass. The TC2 model generally
predicts a number of scalars, and some of them couple very
strongly to the top quark. So, we expect the TC2 corrections to
the top quark properties are larger than those of the other models
which treat generations in an egalitarian manner, such as the
popular minimal supersymmetry model (MSSM)\cite{MSSM}.

Although the various exotic production
processes\cite{top-Higgs,pp-tc,ee-tc,cao1} and the rare decay
modes of the top quark\cite{rare-decay} can serve as a robust
probe of the TC2 model, the role of the dominant decay mode $t \to
W b$ should not be underestimated\cite{Nelson}. One advantage of
this decay mode is that it is free of non-perturbative theoretical
uncertainties\cite{Bigi} and future precision experimental data
can be compared with the accurate theoretical predictions. The
other advantage of this channel is that the $W$-boson, as a decay
product, is strongly polarized and the helicity contents
(transverse-plus $W_+$, transverse-minus $W_-$ and longitudinal
$W_L$) of the $W$-boson can be probed through the measurement of
the shape of the lepton spectrum in the $W$-boson decay\cite{CDF}.
Among the three polarizations of the $W$-boson in the top quark
decaying, the longitudinal mode is of particular interesting since
it is useful to understand the mechanism of EWSB\cite{equivalance}
. Therefore, the study of top quark decaying into the polarized
$W$-boson can provide some additional information about both the
$tWb$ coupling and EWSB. On the experimental side, the CDF
collaboration has already performed the measurement of the
helicity component of the $W$-boson in the top quark decaying from
Run~1 data and obtained the results
\begin{eqnarray}
&&\Gamma_L/\Gamma =0.91 \pm 0.37 (stat.) \pm 0.13 (syst.), \nonumber \\
&&\Gamma_+/\Gamma =0.11 \pm 0.15 \ ,    \nonumber
\end{eqnarray}
where $\Gamma $ is the total decay rate of $t\to Wb$, and
$\Gamma_L $ and $\Gamma_+ $ denote respectively the rates of top
quark decaying into a longitudinal and transverse-plus $W$-boson.
Although the error of these measurements is quite large at the
present time, it is expected to be reduced significantly during
Run~2 of the Tevatron and may reach $1\% \sim 2\%$ at the LHC
\cite{Willenbrock}. On the theoretical side, the predictions of
these quantities in the SM up to one-loop level are now available
\cite{Groot1,Groot2}. The tree-level results are 0.703 for
$\Gamma_L/\Gamma$, 0.297 for $\Gamma_-/\Gamma$ and
${\cal{O}}(10^{-4})$ for $\Gamma_+/\Gamma$, and the QCD
corrections to these predictions are respectively $-1.07 \% $,
$2.19\%$ and $0.10\%$, while the electroweak corrections are at
the level of a few per mill.

In order to probe the new physics from the future precise
measurement of $\Gamma_L/\Gamma$, $\Gamma_-/\Gamma$ or
$\Gamma_+/\Gamma$, we must know the new physics contributions to
these quantities in various models. By now, the one-loop
corrections to the total width of $t \to b W$ in the framework of
MSSM have been studied in \cite{tbw-MSSM}  and the corrections to
$\Gamma_L/\Gamma$, $\Gamma_-/\Gamma$ or $\Gamma_+/\Gamma$ in MSSM
were recently studied in Ref.\cite{cao}, but the similar study in
the TC2 model is absent. Studying the corrections on these
quantities in the TC2 model is the main goal of this paper.

This paper is organized as follows. In the section II, we first
briefly introduce the TC2 model, then we calculate the
corrections and discuss our numerical results. The conclusions are
given in section III.

\section{Top quark decays  into polarized W-boson in the TC2 model}

\subsection{The TC2 Model}

Among various kinds of dynamical electroweak symmetry breaking
models,  the TC2 model \cite{tc2-Hill,tc2-Lane} is especially
attractive since it combines the fancy ideas of
technicolor\cite{technicolor} and top quark
condensation\cite{tc2-2} without conflicting with low energy
experimental data . The basic thought of the TC2 model is to
introduce two strongly interacting sectors. One sector(topcolor
interaction) provides the main part of top quark mass but has the
small contribution to EWSB, while the other sector(technicolor
interaction) is responsible for the bulk of EWSB and the masses of
light fermions. At EWSB scale, this model predicts two groups of
scalars corresponding to the technicolor condensates and topcolor
condensates, respectively\cite{tc2-Hill,tc2-Lane,tc2-2}. Either of
them can be arranged into a $SU(2)$
doublet\cite{2hd,2hd1,Rainwater}, and their roles in TC2 model are
quite analogous to the Higgs fields in the model proposed in
Ref.\cite{special} which is a special two-Higgs-doublet model in
essence. Explicit speaking, the doublet $\Phi_{TC}$ which
corresponds to the technicolor condensates is mainly responsible
for EWSB and light fermion masses, it also contributes a small
portion of top quark mass. Because its vacuum expectation value
(vev) $v_{TC}$ is near the EWSB scale($v_w$), the Yukawa couplings
of this doublet to the fermions are small. While the doublet
$\Phi_{TOPC}$ which corresponds to the topcolor condensates plays
a minor role in EWSB and only couples to the third generation
quarks, its main task is to generate the large top quark mass.
Since the the vev of $\Phi_{TOPC}$ (denoted as $F_t $) can not be
large(see below), the doublet $\Phi_{TOPC}$ can couple strongly to
top quark to generate the expected top quark mass.

One distinct feature of this model is that there exist tree level
flavor changing couplings for the two scalar
fields($\Phi_{TC},\Phi_{TOPC}$)\cite{special} which is
theoretically disfavored. Such defect may be partially alleviated
if the mixing angle between two scalar fields, which is a model
dependent parameter, satisfies $ \tan \alpha =\frac{F_t}{v_{TC}}
$\cite{special}.  In this case, only one scalar field has the
flavor changing couplings and the rearranged Lagrangian has the
following characteristics: one rearranged doublet is fully
responsible for EWSB, but with small Yukawa coupling to all
fermions; while the other rearranged doublet(denoted as: $\Phi$)
has strong Yukawa coupling with the third generation
quarks\cite{cao1}. The Lagrangian relevant to our calculation then
can be written as \footnote{The Lagrangian in Ref.\cite{Rainwater}
corresponds to the case $ \tan \alpha = 0 $. As far as the process
considered in this paper, these two natural choices of $\tan
\alpha $ do not make any significant difference in numerical
results since in both cases, $h_t $ is top condensates dominant.
But our choice of $\tan \alpha $ will make the calculation
simplified.}
\begin{eqnarray}
{\cal L}= | D_{\mu} \Phi |^2 - Y_t
\frac{\sqrt{v_{w}^{2}-F_{t}^{2}}} {v_{w}} \bar{\Psi}_L \Phi t_R -
Y_t \frac{\sqrt{v_{w}^{2}-F_{t}^{2}}} {v_{w}} \bar{t}_R \Phi
\Psi_L  -m_t \bar{t} t  \label{laga}
\end{eqnarray}
where,  $ v_{w} \equiv v/\sqrt{2} \simeq 174$ GeV, $Y_t =
\frac{(1- \epsilon) m_t}{F_t} $ is the Yukawa coupling , $\Psi_L $
is the $SU(2)_L $ top-bottom doublet as usual,  $ \Phi $ is the
rearranged $SU(2) $ doublet and takes the form
\begin{eqnarray}
\Phi =\left ( \begin{array}{c} \frac{1}{\sqrt{2}} ( h_t^0
+ i \pi_t^0 )  \\
\pi_t^- \end{array} \right )
\end{eqnarray}
and the covariant derivative is
\begin{eqnarray}
D_{\mu} = \partial_{\mu}+ i \frac{g_Y}{2} Y B_{\mu} + i
\frac{g}{2} \tau_i W_{\mu}^i
\end{eqnarray}
with the hypercharge of the doublet is $Y =-1 $ and $ g $ is $
g_{weak} $. In Eq.(\ref{laga}), the factor
$\frac{\sqrt{v_{w}^{2}-F_{t}^{2}}} {v_{w}} = \frac{v_{TC}}{v_{w}}$
indicates the mixing effect between the two doublets. The physical
particles ($\pi^0_t,\pi_t^-$) and $h_t^0$ in the $\Phi$ field are
called top-pions and top-Higgs, respectively.

From Eq.(\ref{laga}), one can learn that the TC2 parameters
relevant to our discussion are  $\epsilon$, $F_{t} $ and the
masses of the top-pions and top-Higgs. Before numerical
evaluation, we recapitulate the theoretical and experimental
constraints on these parameters.

In the TC2 model, $\epsilon $ parameterizes the portion of the
extended technicolor contribution to the top quark mass. The bare
value of $\epsilon $ is generated at the ETC scale, and can obtain
a large radiative enhancement from topcolor and $U(1)_{Y_1} $ by a
factor of order $10$ at the weak scale\cite{tc2-Hill}. This
$\epsilon $ can induce a nonzero top-pion mass (proportional to
$\sqrt{\epsilon}$ )\cite{Hill} which can ameliorate the problem of
having dangerously light scalars. Numerical analysis shows that,
with reasonable choice of other input parameters, $\epsilon $ with
order $10^{-2} \sim 10^{-1} $ may induce top-pions as massive as
the top quark\cite{tc2-Hill}. Indirect phenomenological
constraints on $\epsilon $ come from low energy flavor changing
processes such as $ b \to s \gamma $ \cite{b-sgamma}. However,
these constraints are very weak. Precise value of $\epsilon $ may
be obtained by elaborately measuring the coupling strength between
top-pions/top-Higgs and top quark at the next linear colliders.
From theoretical point of view, $\epsilon $ with value from $ 0.01
$ to $ 0.1 $ is favored. For the mode $t\to Wb$ considered in this
paper, $\epsilon $ affects our results via
$Y_t=\frac{(1-\epsilon)m_t}{F_t}$, we fix $ \epsilon =0.1$
conservatively in this paper.

Now, we turn to discuss the parameter $F_t $. The Pagels-Stokar
formula \cite{Pagels} gives the expression of $F_t $ in terms of
the number of quark color $N_c $, the top quark mass $m_t$, and
the scale $\Lambda $ at which the condensation occurs:
\begin{eqnarray}
F_t^2= \frac{N_c}{16 \pi^2} m_t^2 \ln{\frac{\Lambda^2}{m_t^2}}.
\label{ft}
\end{eqnarray}
From this formula, one can infer that, if $t\bar{t} $ condensation
is fully responsible for EWSB, i.e. $F_t \simeq v_w \equiv
v/\sqrt{2} = 174$ GeV, then, $\Lambda $ is about $10^{13} \sim
10^{14} $ GeV. Such a large value is less attractive since one
expects new physics scale should not be far higher than the weak
scale  by the original idea of technicolor
theory\cite{technicolor}. On the other hand, if one believes new
physics exists at TeV scale, i.e., $\Lambda \sim 1 $ TeV, then
$F_t \sim 50$ GeV, which means that $t \bar{t} $ condensation
cannot be wholly responsible for EWSB and the breaking of
electroweak symmetry needs the joint effort of topcolor and other
interactions like technicolor. By the way, Eq.(\ref{ft}) should be
only understood as  a rough guide, and $F_t $ may be somewhat
lower or higher. In this paper, we use the value $F_t =50$ GeV to
illustrate the numerical results.

Finally, we focus on the mass bounds of top-pions and top-Higgs.
On the theoretical side, some estimates have been done. The mass
splitting between the neutral top-pion and the charged top-pions
should be small since such splitting comes only from the
electroweak interactions\cite{mass-pion}. Ref.\cite{tc2-Hill} has
estimated the masses of top-pions using quark loop approximation
and showed that the masses are allowed to be a few hundred GeV in
the reasonable parameter space. Like Eq.(\ref{ft}), such estimates
can only be regarded as a rough guide and the precise values of
top-pion masses can only be determined by future experiments. The
mass of the top-Higgs $h_{t}$ can be estimated in the
Nambu-Jona-Lasinio (NJL) model in the large $N_{c}$ approximation
\cite{NJL} and is found to be about $2m_{t}$
\cite{top-Higgs,2hd1}. This estimates is also rather crude and the
mass below the $\overline{t}t$ threshold is quite possible in a
variety of scenarios \cite{y15}. On the experimental side, the
current experiments have restricted the masses of the charged
top-pions. For example, the absence of $t \to \pi_t^+b$ implies
that $m_{\pi_t^+}
> 165$ GeV \cite{t-bpion} and $R_b$ analysis yields $
m_{\pi_t^+}> 220$ GeV \cite{burdman}. For the masses of neutral
top-pion and top-Higgs, the experimental restrictions on them are
rather weak. The direct search for the neutral top-pion
(top-Higgs) via $ p p \to t \bar{t} \pi_t^0 (h_t) $ \footnote{The
production of a top-pion (top-higgs) associated with a single top
quark at hadron colliders, $p p \to t \pi_t^0 (h_t)$, has an
unobservably small rate since there exists severe cancellation
between diagrams contributing to this process \cite{Rainwater}.}
with $\pi_t^0 (h_t) \to b \bar{b} $ has been proved to be hopeless
at Tevatron with the top-pion (top-Higgs) heavier than $120 $ GeV
\cite{Rainwater}. The single production of $\pi_t^0 $ ($h_t $ ) at
Tevatron with $\pi_t^0 $ ($h_t $) mainly decaying to $t \bar{c} $
may shed some light on detecting the neutral top-pion
(top-Higgs)\cite{top-Higgs}, but the potential for the detection
is limited by the size of the mixing between top and charm quarks.
On the other hand, the detailed background analysis is absent now.
Anyhow, these mass bounds will be greatly tightened at the
upcoming LHC \cite{pp-tc,Rainwater}. In our following discussion,
we will neglect the mass difference among the top-pions and denote
the mass of them as $m_{\pi_t}$.

\subsection{Top quark decays into the polarized W boson}
\begin{figure}[hbt]
\begin{center}
\epsfig{file=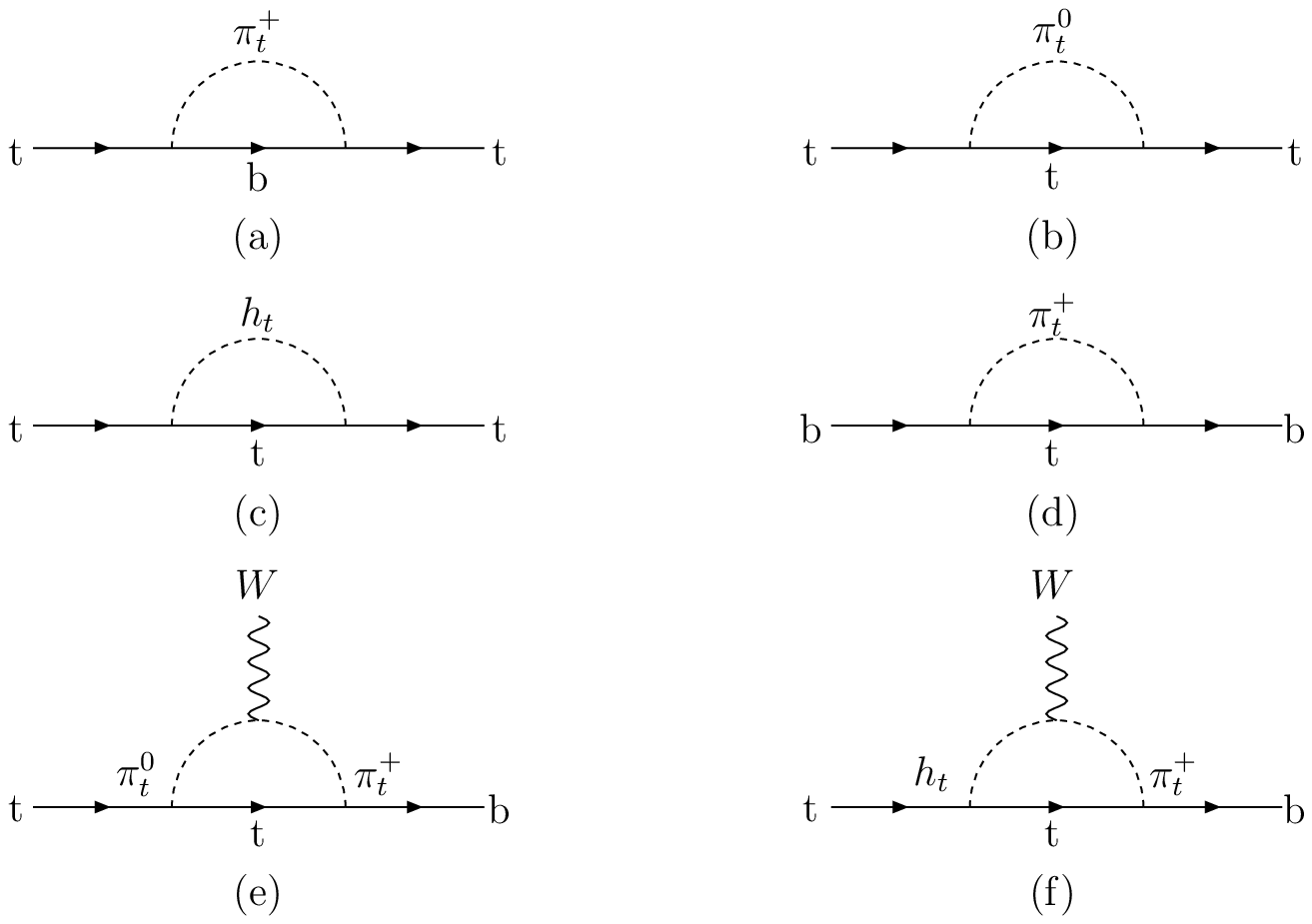,width=11cm}
\caption{The Feynman diagrams for TC2 correction to $t \to b W $.
} \label{feynman}
\end{center}
\end{figure}

Generally speaking, the effective $tbW $ vertex  at one loop level
gets the contribution from penguin diagrams, fermion self-energy
diagrams as well as W boson self-energy diagrams. As far as TC2
model is concerned, the leading part of the first two kinds of
diagrams is $ {\cal{O}}(Y_t^2) $, while that for the last kind of
diagrams is ${\cal{O}} (g^2) $. Considering $Y_t^2 \gg g^2 $, we
can safely neglect the contribution of W boson self-energy. So,
the diagrams we need to calculate are only those shown in
Fig.\ref{feynman}. The effective $tbW $ can be written as
\begin{equation}
\Gamma^{\mu}=-i\frac{g V_{tb}}{\sqrt{2}}\{\gamma^{\mu}P_{L}
[1+F_{L}+\frac{1}{2}\delta Z_{b}^{L}+\frac{1}{2}\delta Z_{t}^{L}
]+\gamma^{\mu}P_{R}F_{R}+P_{t}^{\mu}P_{L}\widetilde{F}_{L}+P_{t}^{\mu}
P_{R}\widetilde{F}_{R}\}
\end{equation}
Here $P_{R,L}\equiv\frac{1}{2} \left ( 1 \pm \gamma_5 \right )$
are the chirality projectors. The form factors $F_{L,R}$ and
$\widetilde{F}_{L,R}$ represent the contributions from the
irreducible vertex loops. $\delta Z_b^L$ and  $\delta Z_t^L$
denote respectively the field renormalization constants for bottom
quark and top quark. The explicit expressions are given by (we
have neglected bottom quark mass)
\begin{eqnarray}
F_L&=& \frac{(1-\epsilon)^2}{16 \pi^2 V_{tb}}
\frac{v_w^2-F_t^2}{v_w^2} \left ( \frac{ m_t}{\sqrt{2} F_t} \right
)^2 (2 C_{24}^e + 2 C_{24}^f )  \label{factor1}  \\
F_R&=& 0 \\
\widetilde{F}_L &= & 0 \\
\widetilde{F}_R &=& \frac{(1-\epsilon)^2}{16 \pi^2 V_{tb}}
\frac{v_w^2-F_t^2}{v_w^2} \left ( \frac{ m_t}{\sqrt{2} F_t} \right
)^2  2 m_t ( C_0^e + 2 C_{11}^e +  C_{21}^e -C_{12}^e -C_{23}^e
\nonumber \\
& & + C_{11}^f + C_{21}^f -C_0^f - C_{11}^f -C_{23}^f -C_{12}^f )
\\
\delta Z_t^L &=&\frac{(1-\epsilon)^2}{16 \pi^2 }
\frac{v_w^2-F_t^2}{v_w^2} \left ( \frac{ m_t}{\sqrt{2} F_t} \right
)^2 [ B_1^b +B_1^c+2 m_t^2 (B_1^{a \prime} +B_1^{b \prime}+B_1^{c
\prime}+ B_0^{b \prime} -B_0^{c \prime})]
\\
\delta Z_b^L &=& \frac{(1-\epsilon)^2}{16 \pi^2 }
\frac{v_w^2-F_t^2}{v_w^2} \left ( \frac{ m_t}{\sqrt{2} F_t} \right
)^2 2 B_1^d  \label{factor2}
\end{eqnarray}
where the functions $B_{0, 1} $ and $ C_{0, i j} $ are
respectively two-point, three-point Feynman integrals defined
in\cite{Axelrod} and their functional dependences are
\begin{eqnarray}
C^e_{0,i j} &= & C_{0, i j} (-p_t, p_w, m_t, m_{\pi_t}, m_{\pi_t}
), \nonumber \\
C^f_{0,i j} &= & C_{0,i j} (-p_t, p_w, m_t, m_{h_t}, m_{\pi_t} ),
\nonumber \\
B^a_{0,1} & = & B_{0,1} (-p_t, m_b, m_{\pi_t} ), \nonumber \\
B^b_{0,1} & = & B_{0,1}(-p_t, m_t, m_{\pi_t} ),  \nonumber \\
B^c_{0,1} & = & B_{0,1}(-p_t, m_t, m_{h_t} ),  \nonumber \\
B^d_{0,1} & = & B_{0,1}(-p_b, m_t, m_{\pi_t} ), \nonumber
\end{eqnarray}
respectively, and $B^\prime_{0,1} $ denotes $\partial B_{0,
1}/\partial p^2 $.

The rate of the top quark decaying into the polarized $W$-boson
can be obtained either by helicity amplitude method\cite{helicity}
or by the project technique introduced in
Ref.\cite{Groot1,Groot2}. Their expressions are given by
\begin{eqnarray}
\Gamma_L =\frac{g^2 m_t |V_{tb} |^2 }{64 \pi } \frac{(1-
x^2)^2}{x^2} &&\left \{  1+  Re ( \delta Z_b^L + \delta Z_t^L  +
2 F_L ) +  Re(\widetilde{F}_R) m_t (1- x^2) \right \},  \label{gammal} \\
\Gamma_-=\frac{g^2 m_t |V_{tb} |^2 }{32 \pi } (1-x^2)^2  & & \left
\{ 1+ Re ( \delta Z_b^L + \delta Z_t^L  + 2 F_L ) \right \} ,
\label{gammam}
\end{eqnarray}
where $\Gamma_L $ ($\Gamma_- $) denotes the rate of the top quark
decaying into the longitudinal (transverse-minus ) $W$-boson and
$x=M_W/m_t $. In deriving  Eqs.(\ref{gammal},\ref{gammam}),  we
have neglected the $b$-quark mass for simplicity which will
produce  an uncertainty of several per mille on $ \Gamma_{L,-}$.
Another consequence of neglecting $m_b$ is $\Gamma_+ =0$ due to
angular momentum conservation \cite{Groot1}. Then, the total decay
rate of $t \to b W$ is obtained by $\Gamma=\Gamma_L+\Gamma_-$. For
convenience, we define the ratios
\begin{eqnarray}
\hat{\Gamma}_{L,-}= \Gamma_{L,-}/\Gamma, \label{def}
\end{eqnarray}
which can be measured in experiments. We present the relative TC2
corrections as: $\delta \hat{\Gamma}_{L,-}/\hat{\Gamma}_{L,-}^0$
with $\delta \hat{\Gamma}_{L,-}$ denoting the TC2 corrections and
$\hat{\Gamma}_{L,-}^0$ denoting the SM predictions.  In our
numerical evaluation, we fixed $m_t =178 $ GeV\cite{nature}, $m_b
=0 $, $M_W = 80.451 $ GeV and $ g_{weak}=0.654 $, and vary
$m_{h_t},m_{\pi_t}$ in experimentally allowed region.

\begin{figure}[hbt]
\begin{center}
\epsfig{file=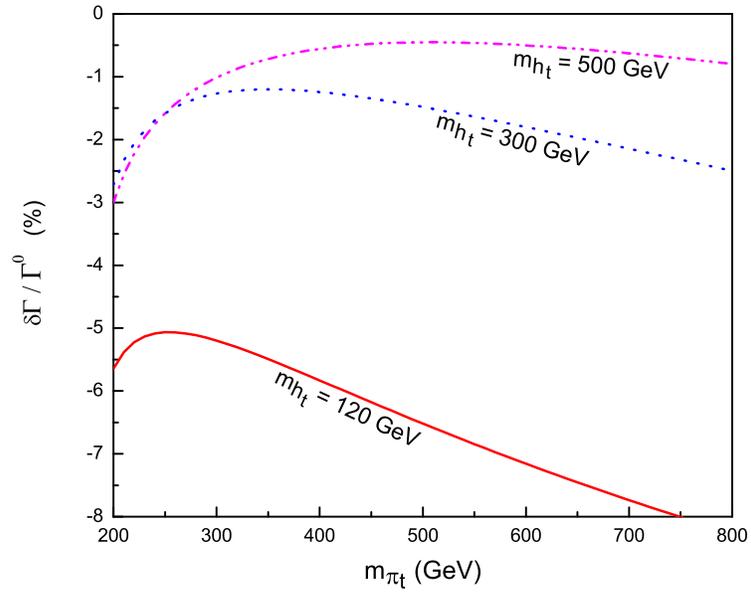,width=11cm}
\caption{The TC2 correction to the width of $\Gamma (t \to b W) $
versus $m_{\pi_t} $ for different $m_{h_t} $.} \label{wid}
\end{center}
\end{figure}

The  Fig.\ref{wid} are the plots of the relative TC2 correction to
the decay width $ \Gamma (t \to W b) $. One distinctive feature of
such correction is that, for fixed $ m_{h_t} $, after the
deviation from the SM predictions reaches its minimum at a certain
value of $m_{\pi_t} $, the relative correction increases
monotonously. This indicates that there are cancellations among
different diagrams\footnote{If different diagram contributions are
constructive, then for fix $m_{h_t} $, the deviation will decrease
monotonously with increasing $m_{\pi_t} $ to approach a
constant.}. Another feature is that the correction is negative in
all allowed parameter space. Noticing the fact that QCD correction
to $ \Gamma (t \to W b) $ is $-8.54 \% $\cite{Groot1,Groot2},  one
can conclude that the TC2 correction can enlarge the quantum
effects. From Fig.2, one can see that, for the light top-Higgs,
the relative correction can reach $ -8\% $. Comparing with the
correction in the popular MSSM model where the SUSY-QCD correction
and the SUSY-EW correction tend to cancel each other\cite{cao}, we
find that the TC2 correction is larger than either of SUSY-QCD and
SUSY-EW correction and TC2 correction might be detectable at the
future high energy colliders\cite{top,review}. In Fig.\ref{wid},
we have fixed $F_t$ as 50 GeV. To get the correction for any other
choice of $F_t $, we just multiply the results of Fig.\ref{wid} by
a factor $ \frac{v_w^2 -F_t^2}{v_w^2 -50^2} \frac{ 50^2}{F_t^2} $
(For example, multiplying a factor 1.6 for $F_t =40 $ GeV and 0.35
for $F_t =80 $ GeV).

\begin{figure}[hbt]
\begin{center}
\epsfig{file=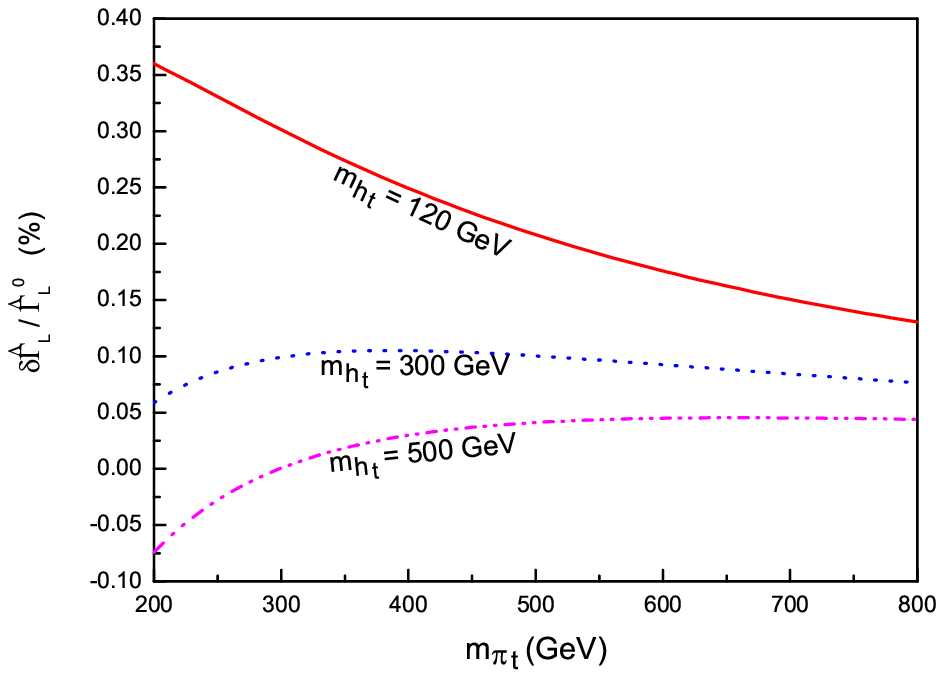,width=11cm}
\caption{The relative TC2 correction to $\hat{\Gamma}_L $ as a
function of $m_{\pi_t} $. } \label{widl}
\end{center}
\end{figure}

\begin{figure}[hbt]
\begin{center}
\epsfig{file=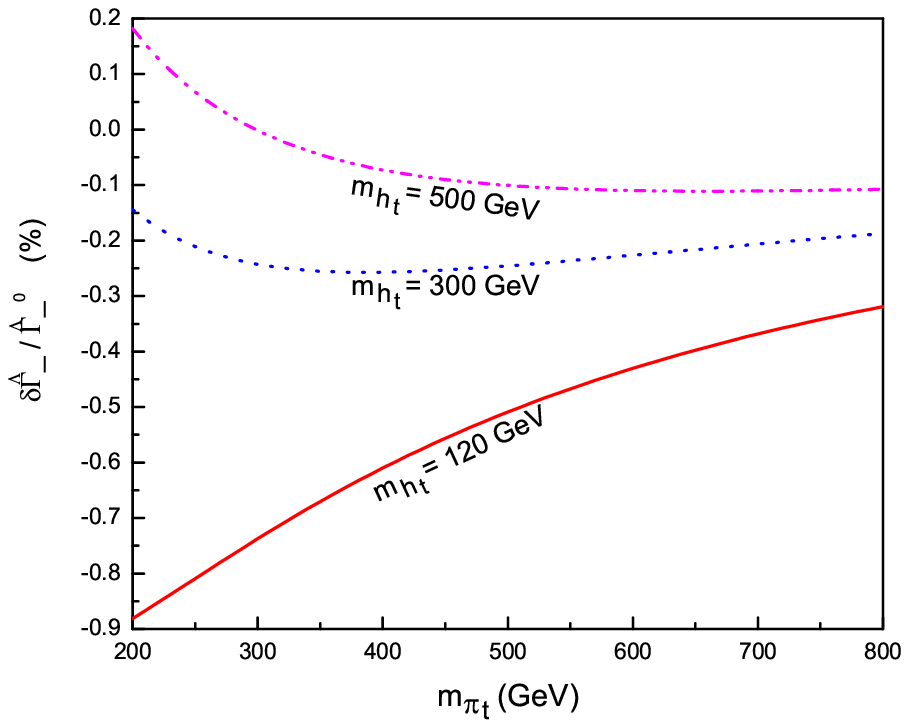,width=11cm}
\caption{The relative TC2 correction to $\hat{\Gamma}_- $ as a
function of $m_{\pi_t} $. } \label{wid-}
\end{center}
\end{figure}

In Fig.\ref{widl} and Fig.\ref{wid-} , we show the relative TC2
correction to $\hat{\Gamma}_L $ and $ \hat{\Gamma}_- $ as a
function of $ m_{\pi_t} $.  One can see that the correction is
below $1 \% $, smaller than the corresponding QCD
correction\cite{Groot1,Groot2} but larger than the corresponding
MSSM corrections\cite{cao}. Comparing with the results in
Fig.\ref{wid}, we can see that the correction to
$\hat{\Gamma}_{L,-} $ is smaller than that to total decay width.
This is due to the cancellation between the correction to $
\Gamma_{L,-} $ and to $\Gamma $ in eq.(14). For example, when we
take $m_{h_t}=120$ GeV and $m_{\pi_t}=750$ GeV, the relative
correction to the total width $\delta\Gamma/\Gamma^0$ is about
$-8\%$($\delta\Gamma_L/\Gamma^0=-5.6\%$ and
$\delta\Gamma_{-}/\Gamma^0=-2.4\%$, respectively), but for the
same values of $m_{h_t}$ and $m_{\pi_t}$,
$\delta\hat{\Gamma}_L/\hat{\Gamma}_L^0=0.13\%$ and
$\delta\hat{\Gamma}_{-}/\hat{\Gamma}_{-}^0=-0.34\%$. Comparing
with the experimental data, we can conclude that the theoretical
prediction of $\hat{\Gamma}_{L,-}$ including the TC2 correction
should be within the experimentally allowed region.

\section{conclusion}

In this paper, we study the TC2 correction to the mode $t \to Wb$.
We find that, due to the cancellations among different diagrams,
the TC2 correction to the width $\Gamma (t \to b W) $ is generally
several percent in the allowed parameter region. The maximum value
of the relative correction can reach $8\% $ which is larger than
that of minimal supersymmetric model and comparable with the QCD
correction. Such TC2 correction should be observable at future
high energy colliders. We also study the TC2 correction to the
branching ratio of top quark decaying into different polarized W
boson states and find the relative TC2 correction is below $1 \%$.

\section*{Acknowledgment}
This work is supported by the National Natural Science Foundation
of China(Grant Nos. 10175017 and 10375017), the Excellent Youth
Foundation of Henan Scientific Committee(Grant No. 02120000300),
and the Henan Innovation Project for University Prominent Research
Talents(Grant No. 2002KYCX009).

\newpage

\newpage

\end{document}